\documentclass[prd,twocolumn,aps,floatfix,superscriptaddress]{revtex4}

\usepackage{amssymb}
\usepackage[dvips]{graphicx}
\usepackage{psfrag}
\usepackage{subfigure}

\usepackage{amsmath,amsfonts}
\usepackage{epsfig}

\def\sfrac#1#2{{\textstyle{\frac{#1}{#2}}}}
\newcommand{\msbar}{\text{$\overline{\text{MS}}$}}

\begin{document}

\title{High-precision determination of the light-quark masses from 
realistic lattice QCD}

\author{Quentin Mason}
\affiliation{Department of Applied Mathematics and Theoretical Physics, 
Cambridge University, Wilberforce Road, Cambridge CB3 0WA, United Kingdom}

\author{Howard D. Trottier}
\affiliation{Simon Fraser University, Department of Physics, 
8888 University Drive, Burnaby, BC, V5A 1S6, Canada}
\thanks{Permanent address.}
\affiliation{TRIUMF, 4004 Wesbrook Mall, Vancouver, BC, V6T 2A3, Canada}

\author{Ron Horgan}
\affiliation{Department of Applied Mathematics and Theoretical Physics, 
Cambridge University, Wilberforce Road, Cambridge CB3 0WA, United Kingdom}

\author{Christine T. H. Davies}
\affiliation{Department of Physics and Astronomy, University of Glasgow,
Glasgow, G12 8QQ, United Kingdom}

\author{G. Peter Lepage}
\affiliation{Laboratory of Elementary-Particle Physics,
Cornell University, Ithaca, New York 14853, USA}

\collaboration{HPQCD Collaboration}
\noaffiliation

\begin{abstract}
Three-flavor lattice QCD simulations and two-loop perturbation theory 
are used to make the most precise determination to date 
of the strange-, up-, and down-quark masses, $m_s$, $m_u$, and $m_d$, respectively. Perturbative matching is required in
order to connect the lattice-regularized bare-quark masses to the 
masses as defined in the \msbar\ scheme, and this is done here
for the first time at next-to-next-to leading (or two-loop) order.
The bare-quark masses required as input come from simulations 
by the MILC collaboration of a highly-efficient formalism (using 
so-called ``staggered'' quarks), with three flavors of light quarks 
in the Dirac sea; these simulations were previously analyzed in a joint 
study by the HPQCD and MILC collaborations, using degenerate $u$ and $d$ 
quarks, with masses as low as $m_s/8$, and two values of the lattice 
spacing, with chiral extrapolation/interpolation to the physical masses. 
With the new perturbation theory presented here, the resulting \msbar\ 
masses are $m^\msbar_s(2~\mbox{GeV}) = 87(0)(4)(4)(0)$~MeV, 
and $\hat m^\msbar(2~\mbox{GeV}) = 3.2(0)(2)(2)(0)$~MeV,
where $\hat m = \sfrac12 (m_u + m_d)$ is the average of the $u$
and $d$ masses. The respective uncertainties are from statistics, 
simulation systematics, perturbation theory, and 
electromagnetic/isospin effects. 
The perturbative errors are about a factor of two smaller than in an 
earlier study using only one-loop perturbation theory.
Using a recent determination of the ratio 
$m_u/m_d = 0.43(0)(1)(0)(8)$ due to the MILC collaboration, these results
also imply $m^\msbar_u(2~\mbox{GeV}) = 1.9(0)(1)(1)(2)$~MeV
and $m^\msbar_d(2~\mbox{GeV}) = 4.4(0)(2)(2)(2)$~MeV.
A technique for estimating the next order in the 
perturbative expansion is also presented, which uses input from simulations 
at more than one lattice spacing; this method is used here in the
estimate of the systematic uncertainties.

\end{abstract}

\maketitle

\section{Introduction}
The strong sector of the standard model contains a number of 
inputs that are \emph{a priori} unknown and must be determined from 
experiment. Knowledge of these fundamental parameters, the quark masses 
and the strong coupling, also requires precise theoretical input,
because of confinement: quarks and gluons cannot be observed as 
isolated particles, hence one can only extract their properties 
by solving QCD for observable quantities such as hadron masses, as 
functions of the quark masses and the strong coupling. Precise values of
the quark masses in particular are valuable in many phenomenological 
applications, such as in placing constraints on new physics beyond 
the standard model. 

Recent breakthroughs in lattice QCD are having a significant
impact on the determination of many hadronic quantities. These 
advances are due to two related developments: the use of perturbation 
theory in the design of so-called improved-lattice discretizations
\cite{Viability,ImprovedStaggered}, which significantly reduces 
cutoff effects; and numerical simulations of a highly-efficient 
formalism (using so-called ``staggered'' quarks), where the correct 
number of light-quark flavors can be included in the Dirac sea, 
and at sufficiently small quark masses \cite{MILCsims}, 
enabling accurate chiral extrapolations to the physical 
region \cite{StaggeredChiral,MILCchiral}.
These developments have eliminated the large and frequently uncontrolled
systematic uncertainties inherent in most other lattice QCD studies,
where the effects of sea quarks are either completely neglected
(working in the so-called ``quenched'' approximation), or 
where the simulations are done with the wrong number of sea quarks 
(and at very heavy masses): comparison of the predictions of
such unrealistic theories with experiment reveals typical 
inconsistencies of 10--20\%, which precludes their use in
many important phenomenological applications.

Recent work by our group, the High Precision QCD (HPQCD) 
collaboration, together with the Fermilab and MILC collaborations, 
has instead utilized unquenched simulations with up, down 
and strange quarks in the Dirac sea. These simulations
are made computationally feasible by using the $O(a^2)$-improved 
staggered-quark action ($a$ is the lattice spacing), with the one-loop 
$O(a^2)$-improved gluon action (hereafter collectively referred to as 
the ``asqtad'' action \cite{ImprovedStaggered,MILCsims}).
We have shown that these three-flavor simulations give results for a wide 
variety of observables that agree with experiment to within systematic
errors of 3\% or less \cite{Confronts}. This precision, and the unified 
description of hadronic physics in many different systems, is a striking 
consequence of having a realistic description of the Dirac sea. 

The application of this program to many other quantities of interest requires 
additional use of perturbation theory, in order to properly match lattice 
discretizations of the relevant operators and couplings onto continuum QCD. 
These perturbative-matching calculations must generally be carried out
at next-to-next-to-leading order (``NNLO'' order, which is generally 
equivalent to two-loop Feynman diagrams), if one is to obtain results 
of a few-percent precision \cite{HReview,QThesis}.
The success of this approach has been demonstrated by a recent
HPQCD determination of the strong coupling at NNLO order,
with the result $\alpha_\msbar(M_z) = 0.1170(12)$ \cite{HPQCDalphas}, 
the most accurate of any method. 

In this paper we apply our perturbation theory program to make
the first-ever NNLO determination of the light-quark masses
(the computationally-simpler case of the additive 
zero-point renormalization for Wilson fermions was previously computed 
at two-loop order in Ref.~\cite{Panagop}, and for static quarks in 
Ref.~\cite{HellerMartinelli}). The quark masses are not 
physically measurable, and as such are only well defined in certain 
renormalization schemes, such as the \msbar\ mass $m^\msbar(\mu)$, 
evaluated at some relevant scale $\mu$. We have done the necessary 
perturbative matching calculations to connect the lattice-regularized 
bare mass $m_0(a)$ for the ``asqtad'' action,
as a function of the lattice spacing, to $m^\msbar(\mu)$.

The lattice spacings and bare quark masses are required as input,
and these have been determined in earlier studies. 
The lattice spacings we use were determined ultimately from 
the $\Upsilon'-\Upsilon$ mass difference \cite{Upsilon},
which is insensitive to the quark masses; the results agree 
within systematic errors of 1.5--3\% with the lattice spacing 
extracted from a wide variety of other physical quantities 
\cite{Confronts,HPQCDalphas}; this eliminates an 
uncontrolled systematic error in studies that are done without the
correct number of sea quarks.

The bare quark masses have previously been determined in a joint
HPQCD and MILC collaboration study \cite{OneLoopMass}, using 
partially-quenched chiral-perturbation theory 
\cite{StaggeredChiral,MILCchiral} to make precise 
extrapolations to the physical region; the simulations used equal 
dynamical $u/d$ quark masses 
as small as $m_s/8$. These bare masses were
used in Ref.~\cite{OneLoopMass} to estimate the \msbar\ masses using 
one-loop perturbation theory \cite{Hein}; the dominant systematic error 
in that determination came from unknown second and higher orders 
in the perturbative matching. Significant progress is made here due 
to our computation of the second-order perturbative matching coefficient.  
We thereby reduce the systematic error from perturbation theory by about 
a factor of two compared to the earlier result of Ref.~\cite{OneLoopMass}. 
The remaining perturbative error is the same size as the current
lattice systematic error, which is largely due to the 
chiral extrapolation/interpolation. 

The staggered-quark formalism had previously been afflicted with several 
poorly-understood problems, which have been tamed with an aggressive 
program of perturbative improvement, leading to the ``asqtad'' action 
that we are using here. With an unimproved-staggered action large 
discretization errors appear, although they are formally only 
$\mathcal{O}(a^2)$ or higher. Many of the renormalization factors 
required to match lattice operators onto continuum quantities are also large
and poorly convergent in perturbation theory for unimproved-staggered
quarks; this is true, for example, of the mass renormalization 
that is needed here. 
It turns out that both problems have the same source, a particular form 
of discretization error dubbed ``taste violation,'' and 
both are ameliorated by use of  the improved-staggered formalism 
\cite{ImprovedStaggered}.
The perturbation theory then shows small renormalizations
\cite{HReview,QThesis,Hein,LeeSharpePT,HQalphaPT}, 
and discretization errors are much reduced \cite{MILCsims}.
Taste violation is strongly probed in certain quantities,
notably the would-be Goldstone meson masses which play an important role
in our analysis, and these effects are taken into account in the 
staggered chiral-perturbation theory \cite{StaggeredChiral}.

A potentially more fundamental concern about staggered fermions relates
to the need to take the fourth root of the quark determinant, in order to
convert the four-fold duplication of ``tastes'' into one quark flavor.
One might imagine that the fourth root would introduce nonlocalities 
(and would induce violations of unitarity), which would prevent
decoupling of the ultraviolet modes of the theory in the continuum limit.
However a great deal of theoretical evidence has been amassed
which demonstrates that the properties of the staggered theory, with the
fourth root, are equivalent to a one-flavor theory, up to the expected
discretization errors \cite{Confronts}; these are again due to 
short-distance taste-changing interactions, which are mediated by  
high-momentum gluons \cite{ImprovedStaggered}.
The locality of the free-field staggered theory is trivial, and is made
manifest in the ``naive'' basis used in Ref.~\cite{ImprovedStaggered}. 
Non-localities do not arise in perturbation theory, since the staggered-quark 
matrix is diagonal in the taste basis, up to those short-distance 
(and calculable) corrections. Most important, it has been shown that 
perturbative improvement of staggered actions correlates exceedingly well with 
nonperturbatively-measured properties of the staggered-fermion matrix, 
providing clear support for the correctness of the fourth-root procedure. 
This includes the measured pattern of low-lying eigenvalues of the 
staggered matrix \cite{Eigenvalues},
and the measured pattern of taste-violating mass differences in the
non-chiral pions \cite{TasteChanging}.

The rest of this paper is organized as follows.
Section~\ref{S:PT} details the computation of the two-loop matching 
factor, using the pole mass as an intermediate matching quantity.
In Sect.~\ref{S:Results} we use the bare-quark masses from the
MILC ``asqtad'' simulations to extract the \msbar\ 
masses, including an analysis of the systematic 
uncertainties. In that connection, we also describe how our perturbation 
theory results can be used to estimate the third order in the 
perturbative expansion, using input from simulations at more than 
one lattice spacing.
Section~\ref{S:Discussion} presents some conclusions, including a
comparison of our results with other recent determinations of the 
strange-quark mass.

\section{Perturbation Theory}\label{S:PT}

\subsection{Lattice to \msbar\ matching}

We do two-loop perturbative matching to connect the cutoff-dependent 
lattice bare-quark masses $m_0(a)$ to the \msbar\ masses $m^\msbar(\mu)$ 
at a given scale $\mu$. We define the perturbative matching factor 
$Z_m(\mu a,m_0 a)$
according to
\begin{equation}
m^\msbar(\mu) = \frac{(a \, m_0)}{a}\; Z_m(\mu a,m_0 a),
\label{e:Zm}
\end{equation}
where we make explicit the fact that the simulation input provides the 
bare masses in lattice units. We compute
$Z_m$ in two stages, using the pole mass $M$ as an intermediate 
matching quantity. We also use our previous determination 
\cite{HQalphaPT,HPQCDalphas} of the relation between the lattice bare 
coupling $\alpha_{\rm lat}$ and the renormalized coupling $\alpha_V(q)$, 
defined by the static potential, to reorganize both sides of the 
matching equation into series in terms of $\alpha_V(q^*)$, at 
an appropriately-determined scale.

In the following subsections we consider in turn the pole mass
in the \msbar- and lattice-regularization schemes. We also provide
some details on the consistent evaluation of the two-loop on-shell
condition in the lattice scheme, and the various checks that we have
applied to our evaluation of the two-loop self-energy. We then
quote our results for the matching factor of Eq.~(\ref{e:Zm}),
and for the relevant matching scale $q^*$.

\subsection{Pole mass in the \msbar\ scheme}
We begin by recalling the relation between the \msbar\ mass
and the pole mass $M$ \cite{Tarrach,Broadhurst}, which is known 
through three loops \cite{Chetyrkin,Melnikov}.
We require it to second order, a result that was first obtained in 
Ref.~\cite{Broadhurst} (expressions for the relation at arbitrary $\mu$
are conveniently given in Ref.~\cite{Chetyrkin})
\begin{align}
   m^\msbar(\mu) = M \Biggl[ 1 +    
   z_1\left(\frac{\mu}{M}\right) \frac{\alpha_\msbar(\mu)}{\pi}  \nonumber\\
   + z_2\left(\frac{\mu}{M}\right) \frac{\alpha^2_\msbar(\mu)}{\pi^2} 
   + \ldots \Biggr],
\label{e:msbarMpole}
\end{align}
where the one- and two-loop coefficient functions
$z_1(\mu/M)$ and $z_2(\mu/M)$ are reduced to a set of terms
with different color structures [in the following 
$C_F = (N_c^2 - 1)/(2 N_c)$, $C_A = N_c$, and $T=1/2$]
\begin{align}
   & z_1 = C_F z_F , \nonumber\\
   & z_2 = C_F^2 z_{FF} + C_F C_A z_{FA} + C_F T n_\ell z_{FL} + C_F T z_{FH} ,
\label{e:z12}
\end{align}
and where the contribution $z_{FH}$ from an internal-quark loop with the
same flavor as the valence quark is split off from the contribution $z_{FL}$
of $n_\ell$ internal-quark loops with different flavor (these are taken here to
be degenerate in mass, though this is trivially generalized). The total number of
flavors is $n_f = n_\ell + 1$. The individual functions are given by
\begin{align}
   &z_F=-1-\sfrac34\ell_{\mu M} ,\nonumber\\
   &z_{FF}=\sfrac{7}{128}-\sfrac{15}{8}\zeta_2-\sfrac34\zeta_3+3\zeta_2\log 2
   +\sfrac{21}{32}\ell_{\mu M}+\sfrac{9}{32}\ell_{\mu M}^{\,2} ,\nonumber\\ 
   &z_{FA}=-\sfrac{1111}{384}+\sfrac12\zeta_2+\sfrac38\zeta_3
   -\sfrac32\zeta_2\log 2-\sfrac{185}{96}\ell_{\mu M}
   -\sfrac{11}{32}\ell_{\mu M}^{\,2} ,\nonumber\\ 
   &z_{FL}=\sfrac{71}{96}+\sfrac12\zeta_2-2\Delta(r_{\rm sea})
   +\sfrac{13}{24}\ell_{\mu M}+\sfrac18\ell_{\mu M}^{\,2}, \nonumber\\
   &z_{FH}=\sfrac{71}{96}+\sfrac12\zeta_2-2\Delta(1)
   +\sfrac{13}{24}\ell_{\mu M}+\sfrac18\ell_{\mu M}^{\,2}, 
\label{e:zFH}
\end{align}
where $\ell_{\mu M} \equiv \log(\mu^2/M^2)$,
\begin{align}
   r_{\rm sea} \equiv \frac{m_{\rm sea}}{m_{\rm valence}} ,
\label{e:rsea}
\end{align}
and where $m_{\rm sea}$ and $m_{\rm valence}$ are the sea- 
and valence-quark masses, respectively. 
The function $\Delta(r)$ gives the dependence of the 
renormalization factors $z_{FL}$  and $z_{FH}$ on the quark mass 
in an internal-quark loop (sea and valence, respectively). 
An exact integral expression for $\Delta(r)$ can be found in 
\cite{Broadhurst}; particular limits are
\begin{align}
   &\Delta(r \ll 1) = \sfrac34 \zeta_2 \, r + O(r^2) , \nonumber\\
   &\Delta(1) = \frac{\pi^2-3}{8} , \nonumber \\ 
   &\Delta(r\gg1) = \sfrac14 \log^2r + \sfrac{13}{24}\log r \nonumber \\
   &\ \ \ \ + \sfrac14 \zeta_2 + \sfrac{151}{288} + O(r^{-2} \log r) . 
\label{e:Delta}
\end{align}

\subsection{Pole mass in the lattice scheme}\label{S:LatticePT}
The relation between the pole mass $M$ and the bare quark mass $m_0(a)$ 
in the lattice-regularized theory has the form 
\begin{align}
   M &= m_0 \Bigl[1 
   +\alpha_{\rm lat} \left(A_{11}\log m_0 a+A_{10}\right) \nonumber \\
   &+\alpha_{\rm lat}^2 \left( A_{22}\log^2m_0a + A_{21}\log m_0 a+A_{20} \right)
   + \ldots \Bigr] ,
\label{e:LatticeMpole}
\end{align}
where $\alpha_{\rm lat}(a)$ is the bare lattice-regularized coupling.
The coefficients of the logarithms are determined by the 
anomalous dimension in $m_0(a)$, which in turn can be determined 
from the known anomalous dimension of the $\msbar$ mass, as described in 
Ref.~\cite{QLattice}. This also requires the connection between
$\alpha_{\rm lat}$ and $\alpha_\msbar$, which can be found from 
our evaluation of the relation between $\alpha_{\rm lat}$ and 
$\alpha_V(q)$, which we obtained at NNLO order in 
Refs.~\cite{HQalphaPT,HPQCDalphas} (see also Ref.~\cite{Schroder}); 
here we only require the NLO relation 
\begin{align}
   \alpha_{\rm lat} = \alpha_V(q)
   \left[ 1 - v_1(q) \alpha_V(q) + \ldots \right] ,
\label{alphaV}
\end{align}
where
\begin{align}
   v_1(q) = \frac{\beta_0}{4\pi} \ln\left( \frac{\pi}{aq} \right)^2
  + v_{10} ,
  \quad \beta_0 = 11 - \frac{2}{3} n_f ,
\end{align}
and where the constant $v_{10}$ for the ``asqtad'' action is
\begin{align}
   v_{10} = 3.57123(17) - 1.196(53)\times 10^{-4} \, n_f .
\end{align}
We then obtain (SU(3) color is used throughout) \cite{QLattice}
\begin{align}
   &A_{11} =  2/\pi, \nonumber \\
   &A_{22} =  0.7599 - 0.03377 \, n_f, \nonumber \\
   &A_{21} = -4.2164 + 0.09429 \, n_f - 0.6366 \, A_{10} .
\end{align}

The terms $A_{10}$ and $A_{20}$ require the computation of the one-
and two-loop lattice self-energies, respectively. We have previously 
determined the first-order term for the ``asqtad'' action 
\cite{Hein,OneLoopMass} 
\begin{equation}
   A_{10} = 0.5432(1) ,
\end{equation}
neglecting corrections of $O((a m_0)^2)$. What is new here is the 
evaluation of $A_{20}$, which is detailed in the next subsection. 

As with the quark-loop terms in the \msbar\ connection to the pole mass,
$A_{20}$ depends on the number of quark flavors in the sea, and on the 
loop-quark masses.
However, as we demonstrate explicitly in the next subsection,
the mass dependence in $A_{20}$ cancels precisely in the matching to 
the continuum relation Eq.~(\ref{e:msbarMpole}), in the limit where the sea 
and valence quark masses are all much less than the lattice ultraviolet 
cutoff $\pi/a$, and the \msbar\ scale $\mu$. The cancellation holds
for an arbitrary ratio $m_{\rm sea}/m_{\rm valence}$ [compare with the 
mass-dependent corrections in the continuum quantities $z_{FL}$ and 
$z_{FH}$ in Eqs. (\ref{e:z12}) and (\ref{e:zFH})].
This follows from the fact that, in the limit where the energy scales
are large, the net-renormalization factor connecting $m_0(a)$ to 
$m^\msbar(\mu)$ probes internal loops at scales large compared to 
the quark masses; hence the 
dependence on quark masses of the intermediate renormalization 
factors connecting to the pole mass $M$ are infrared effects 
that are identical in the lattice- and \msbar-regularized theories.

\begin{figure*}
\begin{center}
\includegraphics[width=5in]{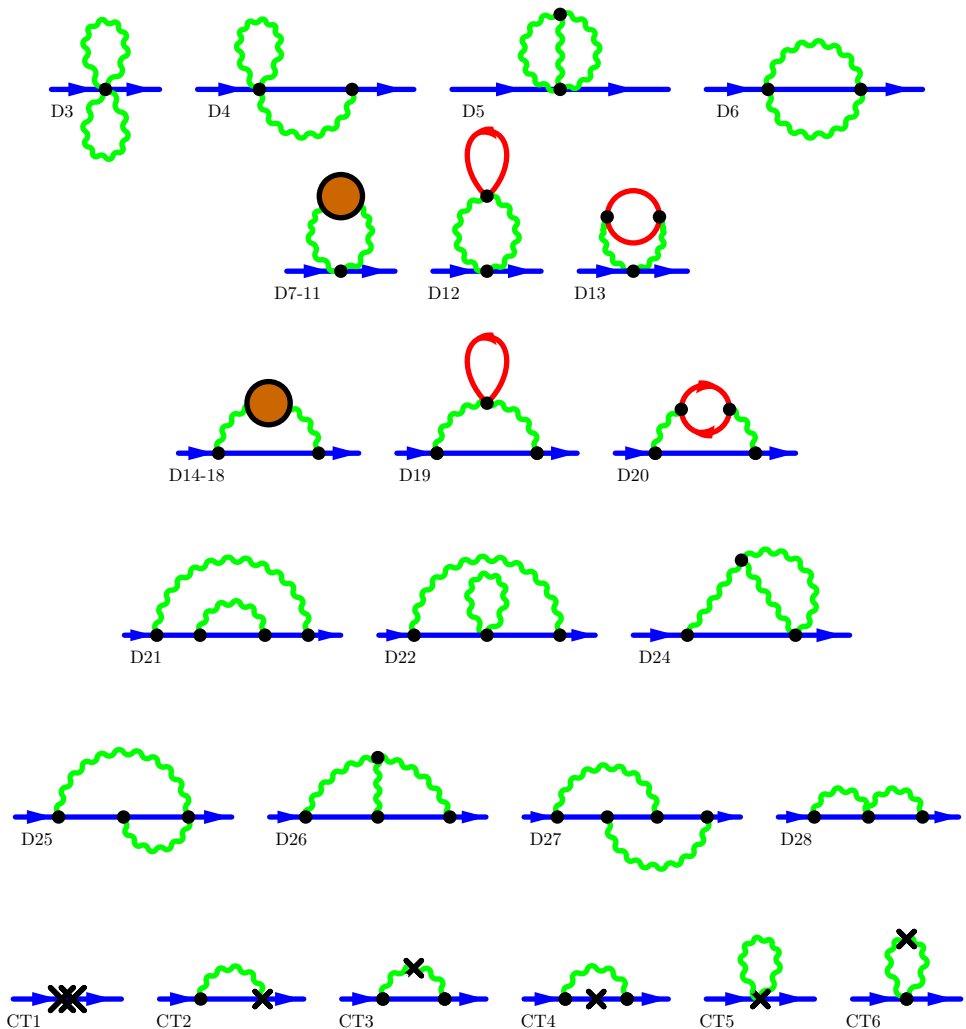}
\caption{\label{f:Diagrams}The two-loop diagrams that contribute to 
the pole-mass renormalization in the lattice scheme.  
The numbering is consistent with 
Ref.~\protect\cite{Panagop}. The filled circles represent five 
one-loop sub-diagrams which dress the gluon propagator at that order
(these contain gluon and ghost loops, and the measure term), in
addition to the diagrams with internal-quark loops, which are shown 
explicitly. The quark-loop diagrams are understood to be summed over all 
flavors. The crosses represent vertices generated
by the perturbative expansion of tadpole and other renormalization 
factors in the gluon and quark actions.}
\end{center}
\end{figure*}

\begin{figure*}
\begin{center}
\includegraphics[width=4.in]{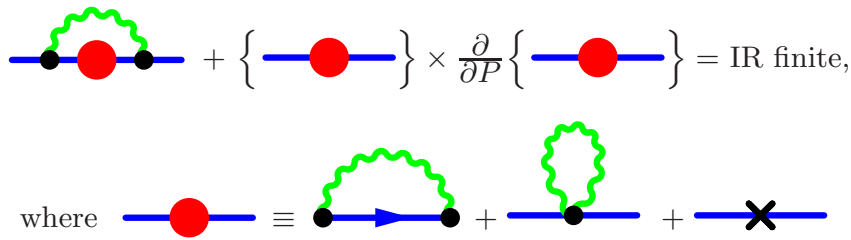}
\caption{\label{f:IRSub}Schematic representation of an IR subtraction
for the two-loop pole mass; appropriate traces of the self-energy 
with an energy projector are implicit.}
\end{center}
\end{figure*}

\begin{figure}[b]
\begin{center}
\includegraphics[width=\columnwidth]{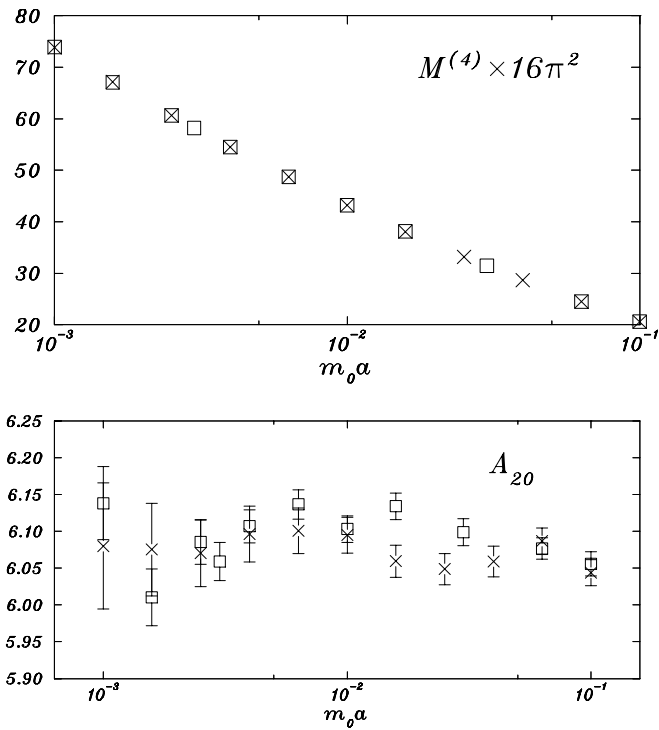}
\caption{\label{f:A20}The $n_f=0$ part of the two-loop 
pole mass $M^{(4)}$ for the ``asqtad'' action (in units of 
$\alpha_{\rm lat}^2$), for varying quark mass $m_0 a$ (upper panel).
The squares and crosses distinguish two independent sets of 
calculations, each one by a different author.
The lower panel shows the results after subtracting 
the known logarithms in $M^{(4)}$, which yields the term $A_{20}$ in 
in Eq.~(\protect\ref{e:LatticeMpole}).}
\end{center}
\end{figure}

\begin{figure}[b]
\begin{center}
\includegraphics[width=\columnwidth]{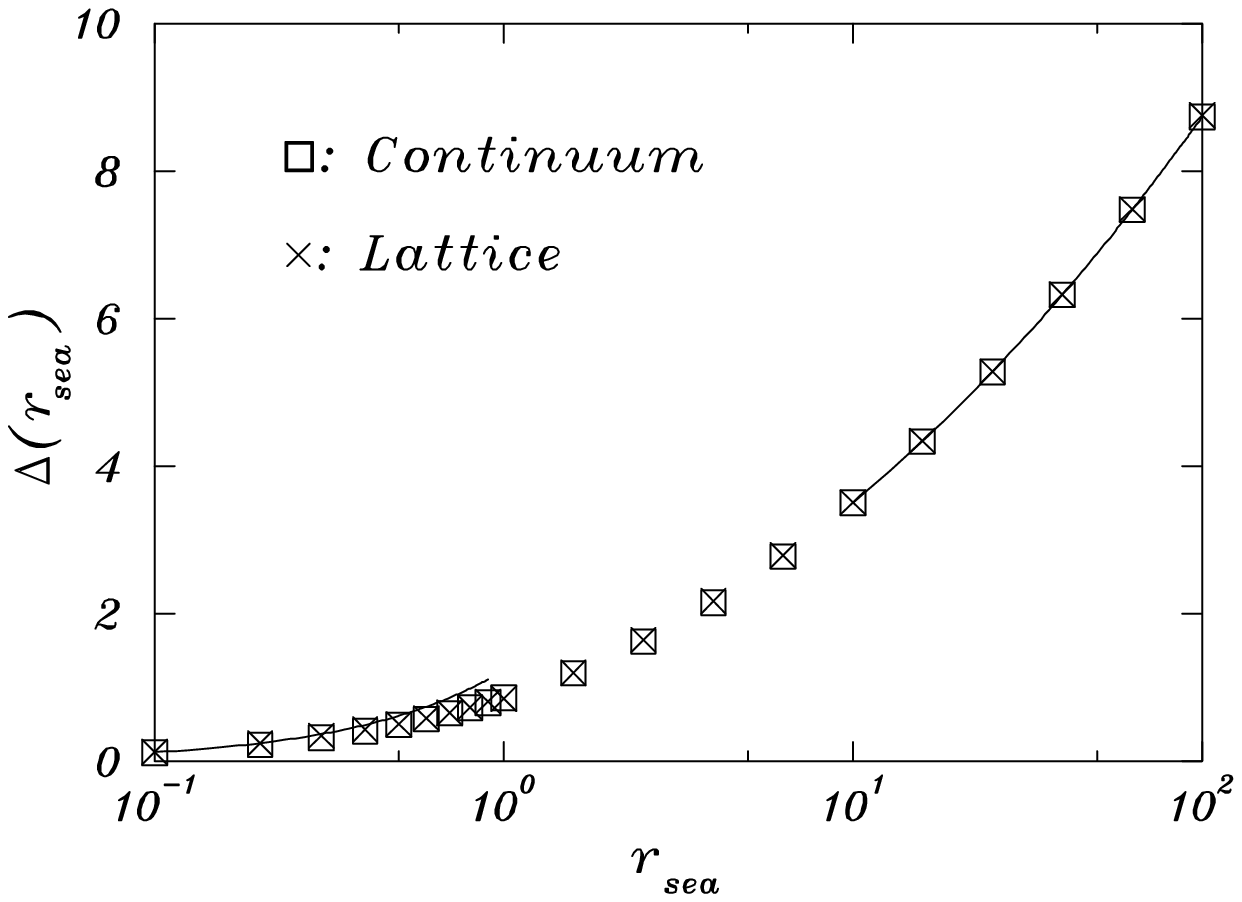}
\caption{\label{f:deltar}Comparison of the loop-quark mass
dependence of pole-mass renormalization factors, on the 
continuum side of the matching, and on the lattice side. 
The lattice calculation used
$am_{\rm valence}=0.001$. The continuum function was taken from
Ref.~\cite{Broadhurst}; the solid lines are from
analytic expressions for the associated limits, cf.\
Eq.~(\protect\ref{e:Delta})}
\end{center}
\end{figure}

\subsection{Evaluation of the two-loop lattice self-energy}
The renormalized pole mass $M$ is identified from the pole in the
dressed-quark propagator $G(p,m_0)$,
\newcommand{\pslash}{\widehat{p}\llap{$\slash$}}
\begin{align}
   G(p,m_0)^{-1} = i\slash\!\!\!\widehat{p} + m_0 + \Sigma_{\rm tot}(p),
\label{e:fullprop}
\end{align}
where $\Sigma_{\rm tot}$ is minus the usual 1PI two-point function.
The two-loop diagrams contributing to $\Sigma_{\rm tot}$ in the lattice
theory are shown in Fig.~\ref{f:Diagrams}.
The lattice-dispersion relation implied by $\widehat p$ will be kept 
as a general function of the lattice momentum $p$; for improved-staggered 
quarks, for instance, 
$a\widehat p_\mu = \sin(ap_\mu) + \frac16 \sin^3(ap_\mu)$. 
We make the spinor decomposition:
\begin{align}
   \Sigma_{\rm tot}(p) = \Sigma_1(p) 
   + (i\slash\!\!\!\widehat{p}+m_0)\,\Sigma_2(p),
\label{e:sigma}
\end{align}
where $\Sigma_1$ and $\Sigma_2$ are both implicit functions of $m_0$,
and both are Dirac scalars. At zero three-momentum the renormalized
on-shell condition is given by
\begin{align}
   P(p_t) \equiv -i\widehat{p}_t = 
   m_0 + \frac{\Sigma_1(p)}{1+\Sigma_2(p)} , 
\label{e:mainsolve}
\end{align}
the solution of which determines the pole mass
\begin{align}
   p_t = iM.
\end{align}

The perturbative expansion of $M$ is denoted by
\begin{align}
   M = M^{(0)}+g_0^2 M^{(2)}+g_0^4 M^{(4)},
\end{align}
with a corresponding notation for other quantities (including, e.g.,
$p_t^{(0)} = i M^{(0)}$, where $P(iM^{(0)}) = m_0$).
At first order one has
\begin{align}
   M^{(2)} = Z_\Psi^{(0)}\,\Sigma_1^{(2)}(p^{(0)}) ,
\label{e:M2}
\end{align}
where
\begin{align}
   {Z_\Psi^{(0)}}^{-1} =\left.\frac{d P(ix)}{dx}\right|_{x=M^{(0)}},
\end{align}
is the tree-level wave function residue. 

An evaluation of the on-shell condition to second order 
requires a consistent expansion of the right-hand side of 
Eq.~(\ref{e:mainsolve}). Part of the $O(g_0^4)$ term 
arises from the one-loop piece of $\Sigma_1(p_t)$, when it is 
evaluated at the  one-loop-corrected on-shell energy
\begin{align}
   \Sigma_{1}^{(2)}(p_t) \simeq
   \Sigma_{1}^{(2)}(iM^{(0)})
   +ig_0^2 \, M^{(2)} \left.
   \frac{\partial \Sigma_1^{(2)}(p_t)}{\partial p_t}\right|_{p_t=iM^{(0)}} .
\end{align}
Hence the two-loop contribution to the pole mass is given by
\begin{align}
   M^{(4)} = Z_\psi^{(0)} \left(m^{(4)} - \sfrac12(M^{(2)})^2 
   {\left.\frac{d^2 P(ix)}{dx^2}\right|_{x=M^{(0)}}} \right) ,
\label{e:M4}
\end{align}
where
\begin{align}
   m^{(4)} = \Sigma_1^{(4)}(iM^{(0)}) +
   \Sigma_1^{(2)}(iM^{(0)}) \,
   {\Sigma_{\rm tot}^{(2)}}^{\,\prime}(iM^{(0)}) ,
\label{e:littlem4}
\end{align}
with  
\begin{align}
   {\Sigma_{\rm tot}^{(2)}}^{\,\prime}(iM^{(0)}) \equiv
   -\Sigma_2^{(2)}(iM^{(0)}) 
   + i Z_\psi^{(0)} \left.\frac{\partial\Sigma_1^{(2)}(p_t)}
   {\partial p_t}\right|_{p_t=iM^{(0)}} .
\label{e:Zlat}
\end{align}
Note that the second-term in Eq.~(\ref{e:M4}) is a correction of 
$O((am_0)^3)$ in the continuum limit of the ``asqtad'' action.

The derivative ${\Sigma_{\rm tot}^{(2)}}^{\,\prime}$ of the one-loop 
self-energy (differentiation is implicitly defined in Eq.~(\ref{e:Zlat}) 
as with respect to $P(p_t)$) 
is also the one-loop part of the wave function renormalization 
(up to finite-lattice discretization corrections), 
and is infrared divergent. This infrared divergence 
precisely cancels against an infrared divergence in the two-loop 
nested-rainbow diagrams (the sum of D21, D22, and CT4 in 
Fig.~\ref{f:Diagrams}), 
which parallels the cancellation of infrared divergences in the 
continuum self-energy at this order. In this connection, we note that 
the continuum-pole mass was first shown to be infrared finite at two loops 
in Ref.~\cite{Tarrach}; an all-orders proof of its finiteness has been 
established only fairly recently, by Kronfeld \cite{Kronfeld}. 

Figure~\ref{f:IRSub} illustrates the infrared cancellation in
Eq.~(\ref{e:littlem4}) in diagrammatic form.
We numerically evaluate the two-loop integral for the left-hand
diagram in the top line of Fig.~\ref{f:IRSub} including, 
in its integrand, a term with the product of the independent integrands 
for the one-loop self-energy, and its derivative.
This grouping is IR finite, and does not require any infrared regulator. 
We obtain a stringent check of this result by noting that this combination
generates a leading-logarithmic contribution to the anomalous dimension 
of the mass which goes like $\log^2(am_0)$, and whose coefficient is 
identical to that of the infrared-subtracted combination in the continuum,
which can easily be read off in Feynman gauge from the \msbar\ results 
in \cite{Tarrach}.  

All the diagrams for the two-loop self-energy, Fig.~\ref{f:Diagrams}, were
generated and evaluated independently by two of us. The Feynman rules for
the highly-improved actions are exceedingly complicated, 
and were generated automatically using computer-algebra 
based codes \cite{HReview,QThesis}.
Moreover, one of us has produced an algorithm which automatically generates
the Feynman diagrams themselves \cite{QThesis}. We can then readily
evaluate the lattice diagrams for a wide variety of gluon and quark 
actions. The two-loop integrals are evaluated numerically, using the 
adaptive Monte-Carlo method {\tt VEGAS} \cite{Vegas}.  
A powerful additional cross-check was provided by an explicit verification
that our results are gauge independent, which we established numerically for 
two different bare-quark masses in three covariant gauges: Feynman, Landau 
and Yennie.

Knowing the coefficients of the logarithmic terms in
Eq.~(\ref{e:LatticeMpole}) also provides a valuable cross-check on 
our results, and increases the accuracy of our numerical determination
of the remaining term $A_{20}$. We compute the two-loop pole mass as a 
function of $m_0 a$, and subtract the known logarithms, in order to isolate 
$A_{20}$, which must be finite as $m_0 a \to 0$. Figure~\ref{f:A20} 
shows the $n_f=0$ part of the results, corresponding to the diagrams without 
internal-fermion loops, over a wide range of bare masses, and clearly 
shows the expected limiting behavior.

Additional checks are provided for diagrams which have a leading
$\log^2(am_0)$ term, which arises from the infrared limit of both 
the outer- and inner-loop integrals, and which is therefore an infrared 
quantity, independent of the ultraviolet regulator; the coefficients
of these double logarithms in the individual diagrams are available in 
Feynman gauge from the original two-loop \msbar\ calculation of 
Tarrach~\cite{Tarrach}, and we have verified that these are reproduced 
in our lattice calculation.   

A further stringent check of our evaluation of the diagrams with
fermion loops (diagrams 12, 13, 19 and 20 in Fig.~\ref{f:Diagrams}) 
is achieved by computing over a range of sea quark masses. 
As discussed in Sect.~\ref{S:LatticePT}, the mass dependence in the
intermediate renormalization from the bare mass to the pole mass $M$
should cancel against the renormalization from $M$ to the \msbar\ mass.
We define
\begin{align}
  A_{20}(r_{\rm sea}) \equiv A_{20}(0) 
+ \frac{4}{3\pi^2} \Delta_{\rm lat}(r_{\rm sea}) ,
\end{align}
and compare with the analogous continuum function $\Delta(r_{\rm sea})$,
cf.\ Eqs. (\ref{e:msbarMpole})--(\ref{e:Delta}). We plot our results in 
Fig.~\ref{f:deltar}, over a very wide range in $r_{\rm sea}$,
which shows the expected agreement.

Our result for the matching term $A_{20}$, in the limit of vanishing 
sea-quark mass, is
\begin{align}
   A_{20} = 6.092(20) - 0.1484(5) n_\ell - 0.0328(3) ,
\label{e:A20}
\end{align}
where the last term corresponds to an internal-quark loop containing the
valence-quark flavor. The uncertainties arise from the numerical evaluation
of the loop integrals. 

The relatively large value of $A_{20}$ may be a symptom of the 
renormalon ambiguity \cite{Renormalon} in the pole mass $M$; there 
is similarly a large second-order term in the
connection between $M$ and the  \msbar\ mass, Eq.~(\ref{e:msbarMpole}). 
However, as seen in the next subsection, these large corrections in the intermediate matchings to the pole mass mostly cancel in the final 
matching of the bare mass to the \msbar\ mass, Eq.~(\ref{e:Zm}), for 
which perturbation theory should be (and is) reliable.

\begin{figure}
\begin{center}
\includegraphics[width=\columnwidth]{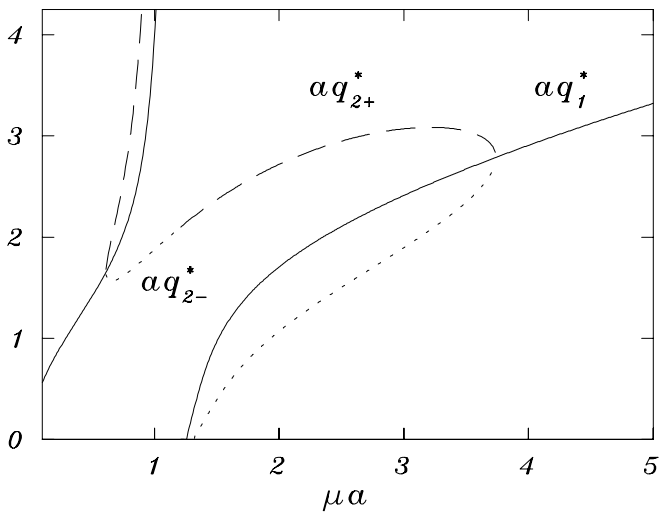}
\caption{\label{f:qstar}BLM scales $aq^*$ for the lattice-bare mass 
to \msbar\ mass matching factor $Z_m(\mu a)$, as functions of $\mu a$. 
The solid line shows the
first-order scale $q_1^*$, while the dashed and dotted lines show the 
second-order scales $q_{2+}^*$ and $q_{2-}^*$, respectively, in regions 
where they are real.
The first-order scale is the correct choice for $\mu a \alt 0.61$ and
$\mu a \agt 3.74$, while the appropriate second-order scale 
applies to the region in between.}
\end{center}
\end{figure}

\subsection{Results for $Z_m$ and its BLM scale}

To complete our determination of the matching factor $Z_m$ in Eq.~(\ref{e:Zm})
we substitute the expression Eq.~(\ref{e:LatticeMpole}) for the pole mass
in the lattice scheme into the equivalent \msbar\ expression in
Eq.~(\ref{e:msbarMpole}), and expand the logarithms in powers of the
coupling. We also reorganize the couplings to the $\alpha_V$ scheme at 
some scale $q^*$, which we determine below according to the 
BLM procedure \cite{BLM}. This leaves an expression with logarithms 
only of $\mu a$ and $aq^*$. The logarithms of $m_0$ drop out of 
$Z_m$, as expected, since these 
are infrared effects that are identical in the intermediate lattice 
and continuum matchings. [We note that the coefficients of the logarithms in
Eq.~(\ref{e:LatticeMpole}), corresponding to the lattice anomalous 
dimension, could instead be determined from the requirement that the 
logarithms of $m_0$ drop out of the final expression for $Z_m$.]

$Z_m$ then takes the form
\begin{align}
   Z_m(\mu a,m_0 a) = 1 + Z_{m,1}(\mu a) \alpha_V\!\left(q^*(\mu a)\right)
   \nonumber \\
   + Z_{m,2}(\mu a)\alpha_V^2 + \mathcal{O}\left(\alpha^3,(m_0 a)^2\right),
\label{e:Zmfinal}
\end{align}
where the first-order term was derived previously in Ref.~\cite{Hein}
\begin{align}
   Z_{m,1}(\mu a) = 0.1188(1) - \frac{2}{\pi} \log(\mu a) .
\end{align} 
The new information presented here is the expression for $Z_{m,2}(\mu a)$, 
which takes the form
\begin{align}
    Z_{m,2}(\mu a) = Z_{22} \log^2(\mu a) + Z_{21} \log(\mu a) + Z_{20},
\end{align}
wherre
\begin{align}
Z_{22} &=  0.7599 - 0.0338 \, n_f , \\
Z_{21} &= -0.4049 - 0.0281 \, n_f \nonumber \\
       &\ + (-1.1145 + 0.0675 \, n_f) \log(aq^*) ,\\
Z_{20} &= 2.086(20) - 0.0144(5) n_f \nonumber \\
       &\ + (0.2080 - 0.0126 \, n_f)\log(aq^*).
\end{align}

To find the optimal choice for $aq^*$ in the BLM scheme, as a function 
of $\mu a$, we analytically evaluated the average-momentum scales in the 
one-loop \msbar\ self-energy diagram (with appropriate renormalizations 
\cite{BLM}), and made numerical evaluations of the average scales in the 
lattice self-energy. We find that the second-order scale-setting procedure of 
Ref.~\cite{BLM} is needed over a wide range of scales in the region 
of $\mu a=1$.

At leading order the scale $q^*$ is determined by \cite{Viability,BLM}
\begin{align}
   \log(q_1^*{}^2) &= \frac{\int d^4 q f(q) \log(q^2)}{\int d^4 q f(q)}
   \equiv \frac{\langle f(q) \log(q^2) \rangle}{\langle f(q) \rangle} ,  
\nonumber \\
   &\equiv \langle\langle \log(q^2) \rangle\rangle
\end{align}
where $f(q)$ is the integrand for the one-loop matching factor,
that is, $\langle f(q) \rangle = Z_{m,1}(\mu a)$. We find
\begin{align}
   \langle f(q) \log(q^2 a^2) \rangle & = 
   -\frac{2}{\pi} \log^2(\mu a) -\frac{5}{3\pi} \log(\mu a) \nonumber \\
   & \ \ -\frac{1}{2\pi} + \frac{5\pi}{12} - 0.821
\end{align}
where the numerical constant is the result of the numerical evaluation of
the lattice moment (logarithms of $m_0$ due to the anomalous dimension 
of the pole mass cancel in the scale for the matching coefficient).

When $\vert Z_{m,1}(\mu a)\vert$ is anomalously small, 
a proper evaluation of $q^*$ requires the second-order expression
\begin{align}
   \log(q_{2\pm}^{*2}) = \langle\langle \log(q^2) \rangle\rangle \pm \left[ 
   \langle\langle \log(q^2) \rangle\rangle^2 - 
   \langle\langle \log^2(q^2) \rangle\rangle \right]^{1/2} ,
\end{align}
where the appropriate root (if the result is real) is usually made obvious
by requiring continuity, and a physically reasonable value, for the 
resulting $q^*$, as a function of the underlying parameters.
The second moment of $Z_{m,1}$ is given by
\begin{align}
   &\langle f(q) \log^2(q^2 a^2) \rangle = 
   -\frac{8}{3\pi}\log^3(\mu a) -\frac{10}{3\pi} \log^2(\mu a)  \nonumber \\
   &\ \ +\frac{(\pi^2 - 9)}{3\pi} \log(\mu a)
   -\frac{5}{3\pi} - \frac{10\,\zeta_3}{3\pi} - \frac{7\pi}{36} + 3.057 .
\end{align}

Results for the first- and (where appropriate) second-order scales, 
as functions of $\mu a$, are shown in Fig.~\ref{f:qstar}. A typical 
value is $aq^* = 1.877$ at $\mu a=1$.

\section{Results for light-quark masses}\label{S:Results}

The bare lattice masses for the strange and up/down quarks, on the
MILC ``coarse'' and ``fine'' lattices, are given in Ref.~\cite{OneLoopMass}.
For the strange quark these are $a m_{0s} = 0.0390(1)(20) / u_{0c}$,
and $a m_{0s} = 0.0272(1)(12) / u_{0f}$, on the coarse and fine
lattices respectively, where $u_{0c}=0.86774$ and $u_{0f}=0.87821$ are tadpole
normalization factors. The uncertainties are lattice statistical
and systematic errors, respectively, the latter
due mainly to chiral extrapolation/interpolation. The lattice spacings 
can be found in Ref.~\cite{HPQCDalphas},
$a^{-1}_{\rm coarse}=1.596(30)$~GeV, and $a^{-1}_{\rm fine}=2.258(32)$~GeV.

Following conventional practice we quote the light-quark \msbar\ masses
at the scale $\mu=2$~GeV, taking three active flavors of quarks ($n_f=3$).
The BLM scales on the two MILC lattices are then
$a q^*_{\rm coarse} = 2.144$ and $a q^*_{\rm fine} = 1.752$ 
\cite{CouplingNote}. This results in two-loop coefficients 
in Eq.~(\ref{e:Zmfinal}) of 
$Z_{m,2}\vert_{\rm coarse} = 1.939(20)$, and 
$Z_{m,2}\vert_{\rm fine} = 2.270(20)$.
We also require the couplings at the relevant scales,
and for this purpose we use the recently determined value 
$\alpha_V^{(n_f=3)}(7.5~\mbox{GeV}) = 0.2082(40)$ \cite{HPQCDalphas}.
We find $\alpha_V(q^*_{\rm coarse}) = 0.2925(92)$, and
$\alpha_V(q^*_{\rm fine}) = 0.2713(76)$ \cite{CouplingNote}.

Putting all of this together, we obtain the following values for the 
\msbar\ strange-quark mass, on the two lattices
\begin{align}
   m^\msbar_s(2~\mbox{GeV}) &= 84(5)~\mbox{MeV\ [coarse]}, \nonumber \\
   m^\msbar_s(2~\mbox{GeV}) &= 86(4)~\mbox{MeV\ [fine]} ,
\label{e:msvalues}
\end{align}
where the errors above are just from the simulation systematics;
we note that these errors are correlated, because the two bare masses
are obtained from a simultaneous-chiral fit to the two lattice spacings, 
which describes the effects of taste-changing and other discretization 
effects in the staggered action \cite{MILCchiral}.

We consider continuum extrapolations of 
these values, based on the form of the expected leading-discretization 
errors, which are of $O(\alpha_V a^2)$ (we find essentially identical 
results by assuming $O(\alpha_V^2 a^2)$ errors \cite{OneLoopMass}).
In addition, we estimate the third-order perturbative correction to
the matching factor, that is, we add a term $Z_{m,3}(\mu a) \alpha_V^3$ 
to the right-hand side of Eq.~(\ref{e:Zmfinal}), and attempt to
estimate its coefficient. To this end, we extend 
the lattice renormalization factor Eq.~(\ref{e:LatticeMpole}) 
to third order, including the logarithms from the three-loop
anomalous dimension, which are fixed from the known \msbar\ 
expansion, along with the known third-order term in Eq.~(\ref{e:msbarMpole}) 
for the \msbar\ mass \cite{Chetyrkin,Melnikov}. This leaves one 
unknown constant, in the lattice renormalization factor, $A_{30}$ in the 
notation of Eq.~(\ref{e:LatticeMpole}), which with the third-order
logarithms determines $Z_{m,3}(\mu a)$. In principle, one can extract
$A_{30}$, and hence the third-order correction, from a simultaneous 
fit using bare lattice masses at several lattice spacings; this extends 
a technique first laid out in Ref.~\cite{HPQCDalphas} (see also
Ref.~\cite{HighBeta}). 

With only the two available lattice spacings, and having also to include 
a discretization correction in the fit, we can only roughly 
bound the size of the next order in the perturbative expansion. 
We used constrained curve-fitting \cite{Constrained,HPQCDalphas} 
to include our expectation that the expansion is convergent
(i.e., $\vert Z_{m,3}(\mu a) \vert = O(1)$ in the notation of
Eq.~(\ref{e:Zmfinal})). 

We tested this procedure by considering a fit to the second-order 
perturbative correction $Z_{m,2}(\mu a)$, without 
{\it a priori\/} knowledge of the associated constant $A_{20}$: with the two 
lattice spacings as input, and including a discretization correction, 
the fit returns $A_{20} = 5.9 \pm 1.9$, in good agreement with 
Eq.~(\ref{e:A20}). The second-order fit also returns 
$m_s^\msbar(2~\mbox{GeV}) = (85 \pm 11)$~MeV, in good agreement with
the ``bona-fide'' two-loop values in Eq.~(\ref{e:msvalues})
[this also represents somewhat of an improvement compared to our
earlier result \cite{OneLoopMass}, which used only {\it a priori\/} 
first-order perturbation theory, without a fit to the second-order
correction, and which somewhat underestimated both
the central value and the systematic error from the 
truncation of the perturbative series].

When this procedure is applied at third-order, with $A_{20}$ input from 
Eq.~(\ref{e:A20}), the fit provides a reasonable estimate of the 
relative systematic error on the \msbar\ mass, due to the third-order 
perturbative correction, of approximately $2 \times \alpha_V^3$, or 
about 4\% (there is no appreciable third-order correction to the 
\msbar\ mass, within this error). We also use these fits (which include
a discretization correction) to extract the central value of the 
strange-quark mass. Our final value is then
\begin{align}
  m^\msbar_s(2~\mbox{GeV}) = 87(0)(4)(4)(0)~\mbox{MeV}
\end{align}
where, following Ref.~\cite{OneLoopMass}, the respective errors are 
statistical, lattice systematic, perturbative, and 
electromagnetic/isospin effects. 

Our result for the ratio of the strange-quark mass to the up/down-quark 
masses is unchanged from Ref.~\cite{OneLoopMass}, since the renormalization 
factor is mass independent, as we have verified explicitly in Sect.~\ref{S:PT}
through two-loops (and up to a negligible mass-dependent discretization 
correction)
\begin{align}
  \frac{m_s}{\hat m} = 27.4(1)(4)(0)(1) ,
\end{align}
where $\hat m \equiv \frac12(m_u + m_d)$. Equivalently we have
\begin{align}
  \hat m^\msbar(2~\mbox{GeV}) = 3.2(0)(2)(2)(0)~\mbox{MeV}
\end{align}
Using a recent determination of the ratio $m_u/m_d = 0.43(0)(1)(0)(8)$ 
due to the MILC collaboration \cite{MILCmud}, these results imply
\begin{align}
   m^\msbar_u(2~\mbox{GeV}) & = 1.9(0)(1)(1)(2)~\mbox{MeV} , \nonumber \\
   m^\msbar_d(2~\mbox{GeV}) & = 4.4(0)(2)(2)(2)~\mbox{MeV} .
\end{align}

\section{Discussion and Conclusions}\label{S:Discussion}

Perturbation theory has once again shown itself to be an 
essential tool in high-precision phenomenological calculations 
from the lattice. The two-loop lattice diagrams for the 
renormalized-quark mass were conquered 
with a combination of algebraic and numerical
techniques in this first-ever two-loop evaluation of a multiplicative 
``kinetic'' mass on the lattice.  When combined with the known 
continuum matching from the pole mass to the
\msbar\ mass, a very accurate determination of the light-quark masses 
was possible.  The results presented here have a number of 
distinguishing features: two-loop perturbation theory; $n_f=2+1$ 
simulations with two degenerate light quarks and a heavier strange quark; 
very small light quark masses from $m_s/8$ to $m_s/2$ which enabled a 
partially quenched chiral fit with many terms to thousands of 
configurations; and extremely accurate determinations of the lattice 
spacings, which are equal within small errors when set from any of
a wide variety of hadronic inputs.

Most notable amongst our results is our new value for strange quark mass,
$m^\msbar_s(2~\mbox{GeV}) = 87(0)(4)(4)(0)~\mbox{MeV}$, where
the respective errors are lattice statistical, lattice systematic 
(mostly due to the chiral extrapolation/interpolation), perturbative, 
and due to electromagnetic/isospin effects. The two-loop matching has 
increased the central value of our estimates of the light-quark masses 
with respect to our previous one-loop determination \cite{OneLoopMass}
by about 1.5 standard deviations, based on the 
previous estimate of the perturbation-theory uncertainty. 
The systematic uncertainty from perturbation theory has been 
reduced by about a factor of two, and is now only about 4\%,
the same size as the current lattice systematic uncertainty, the 
latter due mainly to the chiral extrapolation/interpolation.
We anticipate that the present estimate of the perturbative uncertainty  
could be reduced somewhat further, if additional lattice spacings become 
available, by using the NNLO perturbation theory presented here to 
improve the estimate of the third-order perturbative correction, 
along the lines that we have implemented above. 

The strange-quark mass determination has historically generated some
controversy, with widely different values having been obtained from different
approaches. This is reflected in the large uncertainty in the 
Particle Data Group's most recent best estimate,
$m^\msbar_s(2~\mbox{GeV}) = (105 \pm 25)$~MeV \cite{PDG}; our result 
represents a significant improvement in precision, resulting from an
aggressive effort to understand and reduce all sources of systematic error. 

An obvious advantage of our result is that it has been
obtained with the correct description of the sea, that is, with
$n_f=2+1$ flavors of dynamical quarks. There is only one other 
three-flavor result, which is due to the CP-PACS and
JLQCD collaborations (which did simulations at much larger $u/d$ quark
masses than in the MILC ``asqtad'' ensembles), which recently reported a value
of $m^\msbar_s(2~\mbox{GeV}) = 87(6)$~MeV \cite{CPPACS-JLQCD};
however they do not include a full error analysis, and in particular 
the error from missing higher orders in the perturbative matching alone 
should be comparable to that in our older result, and significantly larger
than the errors that we report here with a 2-loop analysis.

It appears that the most recent estimates of the strange-quark mass 
extracted from simulations with only two flavors of sea quarks are 
systematically higher than the estimates with the correct $n_f$, 
although the other systematic errors are too large to allow for a 
definitive assessment  (noting that these two-flavor determinations
were also done with different definitions of the quark mass and
determinations of the lattice spacing from $r_0$ using different
physical values for $r_0$). The two-flavor 
determination from the QCDSF-UKQCD collaboration 
is $m^\msbar_s(2~\mbox{GeV}) = 100-130$~MeV \cite{QCDSF-UKQCD}, 
the ALPHA collaboration value is $97(22)$~MeV \cite{ALPHA}, 
and the Rome value is
$101(8)\binom{\scriptstyle +25}{\scriptstyle -9}$~MeV \cite{Rome}.
In this connection, we analyzed quenched simulations of the ``asqtad'' 
action by the MILC collaboration \cite{MILCsims,MILCchiral}, and we 
find that this also leads to a somewhat larger value of the strange 
quark mass, of about 96~MeV quenched (using simple linear interpolations 
of the MILC quenched-meson spectrum results, and our two-loop 
perturbative matching formula at $n_f=0$).

We are currently in the process of applying our NNLO matching
calculation to heavy-quark masses, in order to complete a 
high-precision determination of the fundamental parameters 
of QCD. Work is also underway on the NNLO matching 
calculations for important hadronic matrix elements, especially 
those such as the decay constants $f_D$ and $f_B$ and other form-factors
of particular relevance to heavy-flavor physics. 

\acknowledgments
This work was supported by the US Department of Energy, the 
US National Science Foundation, the Natural Science and 
Engineering Research Council of Canada,
the Particle Physics and Astronomy Research Council of the UK, and Barclays
Capital. We thank Matthew Nobes and Kent Hornbostel for fruitful discussions.


\begin{thebibliography}{99}
\bibitem{CouplingNote} The values of the $q^*$ scales used here
are more accurate than the slightly larger values given in
Ref.~\cite{OneLoopMass}, where an approximate determination was 
made of the average-loop momentum circulating in the continuum 
self-energy. In addition an earlier, approximate (and somewhat smaller),
determination of $\alpha_V^{(n_f=3)}(7.5~\mbox{GeV})$ was employed
in Ref.~\cite{OneLoopMass}.
\end{thebibliography}
\end{document}